# Magnetism, upper critical field and thermoelectric power of magneto-superconductor $RuSr_2Eu_{1.5}Ce_{0.5}Cu_2O_{10-\delta}$

R. Lal , V.P.S. Awana[*], and H. Kishan

National Physical Laboratory, Dr. K.S. Krishnan Marg, New Delhi-110012, India.

Rajeev Rawat, and V. Ganesan

UGC-DAE Consortium for Scientific Research, University Campus, Khandwa Road, Indore-452017, India

A.V. Narlikar[$], M. Peurla and R. Laiho

Wihuri Physical Laboratory, University of Turku, FIN - 20014, TURKU, Finland.
[$]Also at UGC-DAE Consortium for Scientific Research, University Campus, Khandwa Road, Indore-452017, MP, India.

Magnetic susceptibility, *M-H* plot, magnetoresistance and thermoelectric power of the $RuSr_2Eu_{1.5}Ce_{0.5}Cu_2O_{10-\delta}$ superconductor are measured. Values of the magnetic transition temperature $T_{mag}$, superconductivity transition temperature $T_c$, upper critical field $H_{c2}$, chemical potential $\mu$, and energy width for electric conduction $W_\sigma$ are obtained from these measurements. It has been found that $T_{mag}$ = 140 K, $T_c$ = 25 K (33 K) from susceptibility (magnetoresistance) measurements, $H_{c2}$ (0) > 32 T, $\mu$ = 8 meV, and $W_\sigma$ = 58.5 meV. These values are compared with other ruthenate superconductors, and resulting physical information is discussed.

**I. Introduction**

Coexistence of superconductivity and magnetism was reported in the ruthenium copper oxide materials $RuSr_2(Gd,Sm,Eu)_{1.6}Ce_{0.4}Cu_2O_{10-\delta}$ (Ru-1222) in 1997 [1,2], and later in $RuSr_2GdCu_2O_8$ (Ru-1212) in 1999 [3-5]. Both of these oxides were synthesized and studied for their transport properties already in 1995 [6]. The Ru-1212 phase is structurally related to the $CuBa_2YCu_2O_{7-\delta}$ (Cu-1212) phase such that the Cu-O chain of Cu-1212 is replaced by the $RuO_2$ sheet. In the Ru-1222 structure furthermore, a three-layer fluorite-type block instead of a single oxygen-free $R$ (= rare earth element) layer, is inserted between the two $CuO_2$ planes of the Cu-1212 structure [7].

Substantial work has been carried out on the Ru-1212 phase. The magnetic structure was studied through neutron diffraction experiments [8]. Electron microscopic works revealed a super-structure along the *a-b* plane due to tilting of the $RuO_6$ octahedra [9], which was further confirmed from neutron diffraction studies [10]. The appearance of bulk superconductivity at low temperatures in Ru-1212 was initially criticised by Chu et. al [11]. However, later works by Bernhard et al [12] and Tokunaga et. al [13] showed that superconductivity exists in this compound within a magnetically ordered state. Couple of substantial review articles/book chapters are also available on these ruthenocuprate magneto-superconductors [14-17].

We notice from the recent work on ruthenocuprates [18-21], that magnetism of the $RuO_2$ layers in Ru-1222 system is quite different from that of Ru-1212. The superstructures due to the tilting of the $RuO_6$ octahedra in Ru-1222 are also qualitatively different than that for the Ru-1212 system. Thus there is a need for investigating the magnetism in these two ruthenocuprates [22]. A related problem is the nature of superconducting and transport behaviour due to the presence of magnetic effect in the ruthenocuprates. Hence we also study this problem. For specificity, we consider the ruthenocuprate $RuSr_2Eu_{1.5}Ce_{0.5}Cu_2O_{10-\delta}$ ($Eu_{1.5}$-1222), and measure its susceptibility, magnetization, magnetoresistance, thermoelectric power and lattice expansion. From the susceptibility measurements we extract the temperature values where different types of magnetic order (antifferromagnetic, weak ferromagnetism, and diamagnetism) take place. The temperature dependence of the upper critical field $H_{c2}$ ($T$) has been extracted from the magnetoresistance data. The thermoelectric power S has been analysed in terms of a normal band model, and the relevant values of parameter are extracted.

Many other workers have also studied the physical properties of different variants of the Ru-1222 system by taking either different rare earth (Gd, Sm etc.) of different content ($Gd_{1.5}$, $Gd_{1.4}$ etc.)

of these ions [18,23,24]. Cardoso et al [18] have made (dc and ac) magnetic measurements on the RuSr$_2$Gd$_{1.5}$Ce$_{0.5}$Cu$_2$O$_{10-\delta}$ (Gd$_{1.5}$-1222) system. They report, in particular a spin glass transition over a significant temperature range. Escote et al [23] have studied three samples of RuSr$_2$Gd$_{1.4}$Ce$_{0.6}$Cu$_2$O$_{10-\delta}$ (Gd$_{1.4}$-1222) with varying oxygen content. They found that for high oxygen content Gd$_{1.4}$-1222 is a superconductor with metallic resistivity in the normal state. When oxygen content is reduced the metallic behaviour of resistivity ρ shrinks to a limited temperature range. They have, in particular studied the occurrence of antiferromagnetic state, and upper critical field. Shi et al [24] have studied the electrical, transport and magnetic properties of RuSr$_2$Sm$_{1.4}$Ce$_{0.6}$Cu$_2$O$_{10-\delta}$ (Sm$_{1.4}$-1222). These authors also find that the metallic portion of the resistivity shrinks with reduced oxygen content. We shall also compare our thermoelectric power data with that of the RuSr$_2$Gd$_{1-x}$La$_x$Cu$_2$O$_8$ sample of Liu et al [25].

It is well known that the physical properties of the ruthenocuprates depend on the preparation conditions [26]. This limits the scope of comparison of our data with that of the data of other groups [18, 23, 24] on different Ru-1222 systems. In fact, an attempt was made in as early as year 2002 on summarising the x-ray-absorption near-edge spectroscopy, electrical resistance, and thermopower measurements on RuSr$_2$Gd$_{2-x}$Ce$_x$Cu$_2$O$_{10+\delta}$ compounds in a commendable way [27]. In particular, a comparison in terms of the specific features of the constituent atoms (like magnetic nature of Gd ions and nonmagnetic nature of Eu ions) will lose its meaning. However, a comparison in terms of the relative values of the physical parameters (ρ, $T_c$, $H_{c2}$ etc.) is expected still to be meaningful. So, below we shall limit to such type of comparison only. We found that this superconductor falls in the clean limit with a mean free path of 56 Å and a coherence length of 24.5 Å. We also estimated the zero temperature upper critical field of Eu$_{1.5}$-Ru-1222 to be $H_{c2}(0)$ = 55 T. A narrow band approach is found to explain the observed thermoelectric power of Eu$_{1.5}$-Ru-1222 above 100 K in terms of two parameters - $\mu$ and $W_\sigma$. The values of $\mu$ and $W_\sigma$ obtained from the fit of the theory with experimental data provides $\mu$ = 8.0 meV and $W_\sigma$ = 58.5 meV.

## II. Experimental

The RuSr$_2$Eu$_{1.5}$Ce$_{0.5}$Cu$_2$O$_{10-\delta}$ sample was synthesized through a solid-state reaction route from RuO$_2$, SrO$_2$, Eu$_2$O$_3$, CeO$_2$ and CuO. Calcinations were carried out on the mixed powder at 1020, 1040 and 1060 $^0$C each for 24 hours with intermediate grindings. The pressed bar-shaped pellets were annealed in a flow of oxygen at 1075 $^0$C for 40 hours and subsequently cooled slowly over a span of another 20 hours down to room temperature. X-ray diffraction (XRD) patterns were obtained using CuK$_\alpha$ radiation. Magnetization measurements were performed on a SQUID magnetometer (Cryogenic Ltd. model S600). Resistivity measurements were made in the temperature range of 5 to 300 K under applied magnetic fields of 0 to 8 Tesla using a four probe technique.

## III. Results and Discussion

### A. x-ray diffraction

Eu$_{1.5}$-1222 copper oxide crystallizes in a tetragonal structure of space group *I4/mmm* with the lattice parameters $a$ = b = 3.8378(2)Å, $c$ = 28.4849(2)Å. An x-ray diffraction pattern for the oxide is shown in Fig. 1. It is to be noted, that a few unidentified lines are also seen in x-ray diffraction pattern; namely at 2θ of nearly 28 and 58 degrees. Though we could not identify them, they are not from Ru-1212, SrRuO$_3$ or another possible culprit RuSr$_2$EuO$_6$. Most probably they arise from the super-structures of the tilted RuO$_6$ octahedras of the system [9, 16, 22]. The lattice parameters and quality of XRD is similar to earlier reported data for various Ru-1222 compounds [1, 2, 14-17]. Ru-1222 compounds are structurally related to the Cu$A_2Q$Cu$_2$O$_{7-\delta}$ [Cu-1$^{(A)}$2$^{(Q)}$12 or Cu-1212, e.g. CuBa$_2$YCu$_2$O$_{7-\delta}$] phase with Cu in the charge reservoir replaced by Ru such that the Cu-O chain is replaced by a RuO$_2$ sheet, furthermore, a three-layer fluorite-type block instead of a single oxygen-free $R$ (= rare earth element) layer is inserted between the two CuO$_2$ planes of the Cu-1212 structure [15-17]. The oxygen content of the present sample is though not determined, it must be in line with our previous works, pl. see ref. 28, in which we showed an oxygen content close to 9.60 for Ru-1222 samples having their $T_c$ close to 30 K and no $T_c$ with oxygent content lower than say 9.40. With this reasoning an oxygen content of nearly 9.60 is assumed for the current sample as its $T_c$ is close to 30 K, to be discussed in coming sections.

### B. Magnetic behavior

Fig. 2 shows the behaviour of the magnetic susceptibility $\chi$ with temperature $T$ in the temperature range of 5 to 300 K for the Eu$_{1.5}$-1222 sample under applied field of 10 Oe. Both types of measurements, the zero-field-cooled (ZFC) and the field-cooled (FC), are shown in this figure. The ZFC and FC curves start branching at 140 K from the higher temperature side with a sharp upward turn at around 100 K. The branching of the FC and ZFC curves signifies onset of the antiferromagnetic (AFM) effect. This means that there is a magnetic transition of the Eu-Ru-1222 system at $T_{mag}$ = 140 K. On moving further towards the low temperature side, it is seen that the ZFC branch shows a cusp at $T_{cusp}$ = 75 K, a superconducting transition temperature at $T_{c,\chi}$ = 25 K, and finally a diamagnetic transition at $T_d$ = 20 K. The down-turn cusp at 75 K in low fields is indicative of the onset of weak ferromagnetism (FM) or spin-glass nature of the Ru spins [18]. This weak FM effect appears due to canted antiferromagnetic spins of the Ru ions. The existence of weak FM below $T_{cusp}$ is seen in the FC branch also. In fact, the FC curve increases fast at $T_{cusp}$, and then saturates at lower temperatures. This shows presence of weak FM in the system.

As compared to $T_{mag}$ = 140 K for the $Eu_{1.5}$-1222 system, the systems $Gd_{1.5}$-1222 [18], $Gd_{1.4}$-1222 [23] and $Sm_{1.4}$-1222 [24] have values of $T_{mag}$ equal to 160, 175 and 150 K respectively. The corresponding values of $T_{c,\chi}$ for these systems are 45, 30 and 28 K, as compared to $T_{c,\chi}$ = 25 K for the $Eu_{1.5}$-1222 system. From these relative values of $T_{mag}$ and $T_{c,\chi}$ we are unable to find whether there is any correlation between these quantities. The only thing we see is that the values of both $T_{mag}$ and $T_{c,\chi}$ are lowest for the present $Eu_{1.5}$-1222 system.

To further elucidate the magnetic property of the $Eu_{1.5}$-1222 superconductor we show isothermal magnetization (*M*) for various values of the applied field (*H*) at 5 K in the inset of Fig. 2. The isothermal magnetization as a function of magnetic field may be viewed as the sum of a linear part and a nonlinear part. That is to say, $M(H) = \chi H + \sigma_s(H)$. Here the linear contribution $\chi H$ arises from the combined effects of the antiferromagnetic (spin-glass) Ru spins and the paramagnetic Eu spins. $\sigma_s(H)$ represents the ferromagnetic component of the Ru moments. The appearance of the $\sigma_s(H)$ at low temperatures within antiferromagnetic/ spin-glass Ru spins could happen due to a slight canting of the spins, as seen from neutron diffraction for another similar magneto-superconductor Ru-1212 [8-10]. The contribution from the weak FM is seen in inset of Fig.2 clearly at 5 K. At *T* > 100 K, the occurrence of weak FM is not so sharp (plot not shown). Combined together, these features are consistence with $T_{cusp}$ = 75 K.

### B. Magnetoresistivity and upper critical field

In Fig. 3 we show the resistivity ($\rho$) of the $Eu_{1.5}$-1222 system up to *T* = 300 K. It is clear that between 75 K and 140 K $\rho$ shows a metallic behaviour. As we have mentioned above, the limited temperature range of metallic behaviour in ruthenocuprate occurs due to decreasing oxygen content [18, 23, 24, 26]. In this sense we expect a linear metallic behaviour at all temperatures (above $T_c$) by the present $Eu_{1.5}$-1222 system if the oxygen content is increased in it. Since the temperature range for the metallic behaviour in the present case, 75 K to 140 K, is a significant range, we do not expect much effect either of the low -$T$ value of $\rho$ or the remaining behaviour of high $T$ on the middle linear portion of $\rho$. When this is so, we may fit the metallic portion of $\rho$ by a straight line (cf. Fig. 3). Extrapolation of this straight (marked in Fig.3) line to $T$ =0 K leads to the zero temperature resistivity $\rho_0$ = 7.9 mΩ-cm. Assuming that the plasma frequency and Fermi velocity of Eu$_{1.5}$-1222 sample have the same values as for the RuSr$_2$Gd$_{1.4}$Ce$_{0.6}$Cu$_2$O$_{10-\delta}$ system [23], and using Eq.(2){$l = 4.95 \times 10^{-4} v_F/(\hbar\omega_p)^2\rho$), $v_F$ is the Fermi velocity, $\hbar$ Planck constant and $\omega_p$ is the plasma frequency} of Ref. [23], we estimate the mean free path of the Eu-Ru-1222 system to be $l = 56$ Å at $T = 0$ K. This is in between the 95-atm and 95-atm – 2x samples of the RuSr$_2$Gd$_{1.4}$Ce$_{0.6}$Cu$_2$O$_{10-\delta}$ system of Ref. [23]. In fact, the 95-atm (95-atm-2x) sample of Ref. [23] corresponds to $l \sim 4$ Å (200 Å). From Fig. 3 we observe that below 75 K the resistivity increases with decreasing temperature. Since above 75 K (but up to 140 K) $\rho$ shows a metallic behaviour, we may argue that the upturn of $\rho$ below 75 K is due to weak localization.

In Fig. 4 we show the magnetoresistance of our sample for the magnetic field values of 0, 1, 2, 4 and 8 T in the temperature range of 0 to 40 K. From these values we estimate $T_c$ ($H$) for all the values of the magnetic field $H$ from the intersection of the top of the transition line and the straight line passing through the linear portion of the $\rho$ - $T$ curve around the point of inflexion near $T_c$. This method is quite familiar for estimating the temperature $T_c$ ($H$) for a given magnetic field in various cuprate superconductors [29]. On the basis of this method we find that $T_c$ ($H = 0$) = 33 K. This is significantly higher than $T_{c\chi}$ = 25 K. However at the same time $T_c$ ($H = 0$) is significantly lower than the onset temperature of 43 K (for $H = 0$) obtained from the $\rho$ versus $T$ plot (Fig. 3). In the following description we shall treat $T_c$ ($H$) as the superconducting transition temperatures of the considered ruthenocuprate for different values of $H$.

The H$_{c2}$ vs $T$ curve, obtained in the above way, is shown in Fig. 5. First of all we see from Fig. 4 that $H_{c2}$ ($T$) has a positive curvature in the observed range of temperatures. This agrees qualitatively with the $H_{c2}$ ($T$) vs $T$ behaviour found for the RuSr$_2$Gd$_{1.4}$Ce$_{0.6}$Cu$_2$O$_{10-\delta}$ superconductor by Escote et al [23]. It may be noted that positive curvature of $H_{c2}(T)$ is an essential feature of the cuprates, and is observed in these systems even at $T<$ 1 K [29,30]. In this sense it becomes imperative to consider such a $T$–dependence of $H_{c2}(T)$ which leads to a positive curvature in the entire temperature range ($T = 0$ to $T = T_c$). In particular the form $H_{c2}$ ($T$) = $H_{c2}(0)$ $[1-(T/T_c)^2]^\alpha$, considered by Escote et. al [23], does not lead to positive curvature near $T = 0$ K, and so we feel that the value of $H_{c2}(0)$ obtained by Escote et al [23] is not reliable. The point is that even at zero curvature, the value of $H_{c2}(0)$ from the estimates of $T_c$ and the slope: $s(T_c) = -dH_{c2}/dT|_{T=Tc}$ turns out to be 44 T. Thus for a positive curvature $H_{c2}(0)$ will essentially be larger than 44 T.

In the present case we have very few points of the $H_{c2}(T)$ versus $T$ relationship. So, it is not possible to obtain a reliable value of $H_{c2}(0)$. However, since the $H_{c2}(T)$ versus $T$ curve should correspond to a positive curvature, the straight line extrapolation up to $T = 0$ of the slope $s(T_c)$ will give a lower limit of $H_{c2}(0)$ is equal to 32 T. This is considerably lower than the corresponding value of 44 T for the $Gd_{1.4}$-1222 system [23]. Using the relation $H_{c2}(0) = \Phi_0/2\pi\xi(0)^2$, where $\Phi_0$ is flux quantum and $\xi(0)$ is zero temperature Ginzburg-Landau coherence length, we find $\xi(0)$ will be less than 32 Å. This value of $\xi(0)$ is much smaller than the mean free path, $l = 56$ Å estimated above. This means that the present sample of $RuSr_2Eu_{1.5}Ce_{0.5}Cu_2O_{10-\delta}$ is in the clean limit. In fact the positive curvature of $H_{c2}(T)$ near $T_c$ also indicates that the superconductor is in the clean limit.

### C. Thermoelectric Power

We show the experimentally observed thermoelectric power $S$ in Fig. 5 by solid squares. We present an analysis of this thermoelectric power on the basis of the equation by Gasumyants et al [31]. This approach for calculating the transport properties of cuprate systems is based on the narrow band picture. Gasumyants et al have obtained an expression of $S$ which is expressed in terms of three parameters – band filling $F$, total effective band width $W_D$ and the effective width of the energy interval $W_\sigma$ for electron conduction. However, when we limit our study to the temperature range given by $2 k_B T \ll W_D$, the approach of Gasumyants *et al* may be expressed only in terms of two parameters – the chemical potential $\mu$ and the effective width $W_\sigma$. Since typically $W_D > 100$ meV [31] and in our measurements $2k_B T < 45$ meV, we can use a two-parameter ($\mu$ and $W_\sigma$) version of the approach of Gasumyants et al. In this sense, we may rewrite Eq. (26) of Gasumyants et al as,

$$S = \left[\frac{k_B}{e\sinh W_\sigma^*}\right]\left[W_\sigma^* \sinh\mu^* + \mu^*\left[\cosh\mu^* + \exp(W_\sigma^*)\right] + \left[\cosh\mu^* + \cosh W_\sigma^*\right]\ln\left\{\frac{1+\exp(W_\sigma^* - \mu^*)}{1+\exp(W_\sigma^* + \mu^*)}\right\}\right] \quad (1)$$

Here, $\mu^* = \mu/k_B T$ and $W^*_\sigma = W_\sigma/k_B T$. Although mathematically Eq. (1) is equivalent to Eq. (26) of Gasumyants *et al*, it (Eq. 1) clarifies that $S$ is an odd function of $\mu$ i.e., $S(-\mu) = -S(\mu)$. This property of $S$ is not obvious from Eq. (26) of Gasumyants et al.

We find that Eq. (1) fits the experimental data very well (Fig.5) for $\mu = 8.0$ meV and $W_\sigma = 58.5$ meV, except below 100 K. The deviation of Eq. (1) from the observed values below 100 K may be attributed to fluctuation effects. For comparison we have also fitted the thermoelectric power data of Liu et al. [25] for the $RuSr_2GdCu_2O_8$ superconductor on the basis of Eq. (1). It is found that $\mu = 28.0$ meV and $W_\sigma = 77.0$ meV give an excellent fitting. This shows that the ratio $W_\sigma/\mu$ is larger in the present case ($W_\sigma/\mu = 7.31$) than that in the case of Liu et al ($W_\sigma/\mu = 2.75$). This, according to Ref. [31], means that the resistivity of the sample of Liu et al. must be significantly larger than that of our sample. This is indeed the case, as the (weakly) metallic properties of the $RuSr_2GdCu_2O_8$ sample of Liu et al corresponds to $\rho_0 = 44$ mΩ-cm (or to $l = 10$ Å), as compared to $\rho_0 = 7.9$ mΩ-cm for the present case of the $Eu_{1.5}$-1222 system. On this basis we may argue that the values of $\mu$ and $W_\sigma$ estimated here provide a consistent understanding of the behaviour of the thermoelectric power.

**IV. Conclusions**

We have synthesized the $Eu_{1.5}$-1222 system, and have measured its various properties. The susceptibility measurements show magnetic order in this system below 140 K, and a superconducting transition at 25 K. The resistivity of the $Eu_{1.5}$-1222 sample shows a metallic behaviour between 75 K and 140 K. From this we estimate a mean free path of 56 Å at zero temperature. The upper critical field of the $Eu_{1.5}$-1222 system shows a positive curvature, and corresponds to a minimum value of the upper critical field equal to 32 T at $T = 0$ K. This value is found to be lower than the corresponding value of the $Gd_{1.4}$-1222 system of Ref. [23]. This is in accordance with the expectation since the $T_c$ of $Eu_{1.5}$-1222 system is alower than that of $Gd_{1.4}$-1222 system. A narrow band approach is found to explain the observed thermoelectric power of $Eu_{1.5}$-1222 above 100 K in terms of two parameters - $\mu$ and $W_\sigma$. The values of $\mu$ and $W_\sigma$ obtained from the fit of the theory with experimental data provides $\mu = 8.0$ meV and $W_\sigma = 58.5$ meV.

**Acknowledgement**

Authors from the NPL appreciate the interest and advice of Professor Vikram Kumar (Director) in the present work. One of us (AVN) thanks University of Turku for providing research facilities and for the invitation for the present visit.

**Figure captions**

Fig. 1. Room temperature XRD patterns for $Eu_{1.5}$-1222 system.

Fig. 2. Behaviour of magnetic susceptibility $\chi$ versus temperature $T$ in the temperature range 5 - 300 K for $Eu_{1.5}$-1222. The inset shows the *M-H* curve at 5 K, 50 K and 100 K for the same.

Fig. 3. Behaviour of Resistivity $\rho$ versus temperature $T$ for $Eu_{1.5}$-1222 up to $T = 300$ K.

Fig. 4. Magnetoresistance of $Eu_{1.5}$-1222 sample for the magnetic field values of 0, 1, 2, 4 and 8 T in the temperature range of 5 to 40 K

Fig. 5. $H_{c2}$ versus $T$ plot for $Eu_{1.5}$-1222 compound.

Fig. 6. Plot of thermoelectric power $S$ versus temperature $T$ for $Eu_{1.5}$-1222 sample. The solid line shows the fitted curve to Eq. (1).

**REFRENCES**

1. I. Felner, U. Asaf, Y. Levi, and O. Millo, Phys. Rev. B 55, R3374 (1997)
2. I. Felner, and U. Asaf, Int. J. Mod. Phys. B. 12, 3220 (1998)
3. C. Bernhard, J.L. Tallon, Ch. Niedermayer, Th. Blasius, A. Golnik, E. Brücher, R.K. Kremer, D.R. Noakes, C.E. Stronack, and E.J. Asnaldo, Phys. Rev. B 59, 14099 (1999).
4. J.L. Tallon, C. Bernhard, M.E. Bowden, T.M. Soto, B. Walker, P.W. Gilberd, M.R. 5. Preseland, J.P. Attfield, A.C. McLaughlin, and A.N. Fitch, IEEE, J. Appl. Supercond. 9, 1696 (1999).
5. D.J. Pingle, J.L. Tallon, B.G. Walker, and H.J. Tordhal, Phys. Rev. B 59, R11679 (1999).
6. L. Bauernfeind, W. Widder and H.F. Braun, Physica C 254, 151 (1995).
7. N. Sakai, T. Maeda, H. Yamauchi and S. Tanaka, Physica C 212, 75 (1993).
8. J.W. Lynn, B. Keimer, C. Ulrich, C. Bernhard, and J.L. Tallon, Phys. Rev. B 61, R14964 (2000).
9. A.C. McLaughlin, W. Zhou, J.P. Attfield, A.N. Fitch, and J.L. Tallon, Phys. Rev. B 60, 7512 (1999).
10. O. Chmaissem, J.D. Jorgensen, H. Shaked, P. Dollar, and J.L. Tallon, Phys. Rev. B 61, 6401 (2000).
11. C.W. Chu, Y.Y. Xue, S. Tsui, J. Cmaidalka, A.K. Heilman, B. Lorenz, and R.L. Meng, Physica C 335, 231 (2000).
12. C. Bernhard, J.L. Tallon, E. Brücher, and R.K. Kremer, Phys. Rev. B 61, R14960 (2000).
13. Y. Tokunaga, H. Kotegawa, K. Ishida, Y. Kitaoka, H. Takigawa, and J. Akimitsu, Phys. Rev. Lett. 86, 5767 (2001).
14. Hans F. Braun "A phase diagram approach to magnetic superconductors" Frontiers in Superconducting Materials: P. 365-392, Ed. by A.V. Narlikar, Springer-Verlag publishing, Germany (2005).
15. I. Felner "Coexistence of superconductivity and magnetism in $R_{2-x}Ce_xRuSr_2Cu_2O_{10}$ (R = Eu and Gd)" Studies of high Temperature Superconductors: P. 41-75, Vol. 46, Ed. by A.V. Narlikar, NOVA Science Publishers, U.S.A (2003): cond-mat/0211533.
16. V.P.S. Awana "Magneto-superconductivity of rutheno-cuprates" Frontiers in Magnetic Materials: P. 531-570, Ed. By A.V. Narlikar, Springer-Verlag publishing, Germany, (2005).

17. C.W. Chu, B. Lorenz, R.L. Meng and Y.Y. Xue, "Rutheno-cuprates: The superconducting ferromagnets" Frontiers in Superconducting Materials: P. 331-364, Ed. by A.V. Narlikar, Springer-Verlag publishing, Germany (2005).
18. C.A. Cardoso, F.M. Araujo-Moreira, V.P.S. Awana, E. Takayama-Muromachi, O.F. de Lima, H. Yamauchi, and M. Karppinen, Phys. Rev. B 67, 020407 (2003).
19. C.S. Knee, B.D. Reinford, and M.T. Weller, J. Mater. Chem. 10, 2445 (2000).
20. I. Zivkovic, Y. Hirai, B.H. Frazer, M. Prester, D. Drobac, D. Ariosa, H. Berger, D. Pavuna, G. Margaritondo, I. Felner, and M. Onillion, Phys. Rev. B, 65, 144420 (2001).
21. C.A. Cardoso, F.M. Araujo-Moreira, V.P.S. Awana, H. Kishan, E. Takayama-Muromachi, O.F. de Lima, Physica C 405, 212 (2004).
22. Tadahiro Yokosawa,, Veer Pal Singh Awana, Koji Kimoto, Eiji Takayama-Muromachi, Maarit Karppinen, Hisao Yamauchi and Yoshio Matsui, Ultramicroscopy 98, 283-295 (2004).
23. M. T. Escote, V.A. Meza, R.F. Jardim, L. Ben-Dor, M.S. Torikachvili and A.H. Lacerda, Phys. Rev. B 66, 144503 (2002).
24. L. Shi, Q. Li, X.J. Fan, S.J. Feng, and X .-G Li, Physica C 399, 64 (2003).
25. C. –J Liu, C. –S. Shew, T.-W. Wu, L. –C. Huang, F. H. Hsu, H.D. Yang, G.V.M. Williams and Chia-Jung C. Liu, Phys. Rev. B 71, 014502 (2005).
26. C. A. Cardoso, A. J. C. Lanfredi, A. J. Chiquito, F. M. Araujo-Moreira, V.P.S. Awana , H. Kishan, and O.F. de Lima, Phys. Rev. B. 71, 134509 (2005).
27. G.V.M. Williams, L. –Y. Jang, and R.S. Liu, Phys. Rev. B 65, 64508 (2002).
28. M. Matvejeff, V.P.S. Awana, H. Yamauchi and M. Karppinen Physica C 392-396, 87 (2003).
29. M.S. Osofsky, R. J. Soulen, Jr., S. A. Wolf, J. M. Broto, H. Rakoto, J. C. Ousset, G. Coffe, S. Askenazy, P. Pari, I. Bozovic, J. N. Eckstein, and G. F. Virshup et al., Phys. Rev. Lett. 71, 2315 (1993).
30. A.P. Mackenzie, S. R. Julian and G. G. Lonzarich, A. Carrington, S. D. Hughes, R. S. Liu, and D. S. Sinclair Phys. Rev. Lett. 71, 1238 (1993); A.S. Alexendrov, N. Zavaritsky, W. Y. Liang, and P. L. Nevsky Phys. Rev. Lett. 76, 983 (1996).
31. V.E. Gasumyants, V.I. Kaidanov and E.V. Vladimirskaya Physica C 248, 255 (1995); V.E. Gasumyauts, N.V. Ageev, E.V. Vladimirskaya, V.I. Smirnov, and A.V. Kazanskiy ,V.I. Kaydanov Phys. Rev. B 53, 905 (1996).

Fig. 1 R. Lal et al.

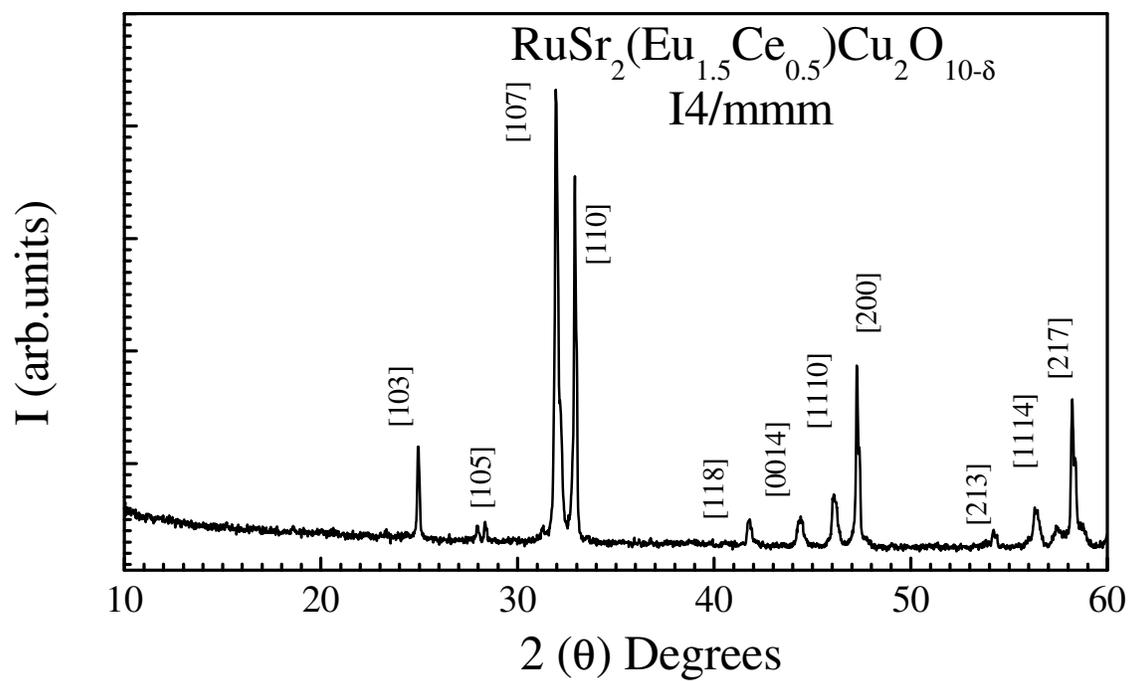

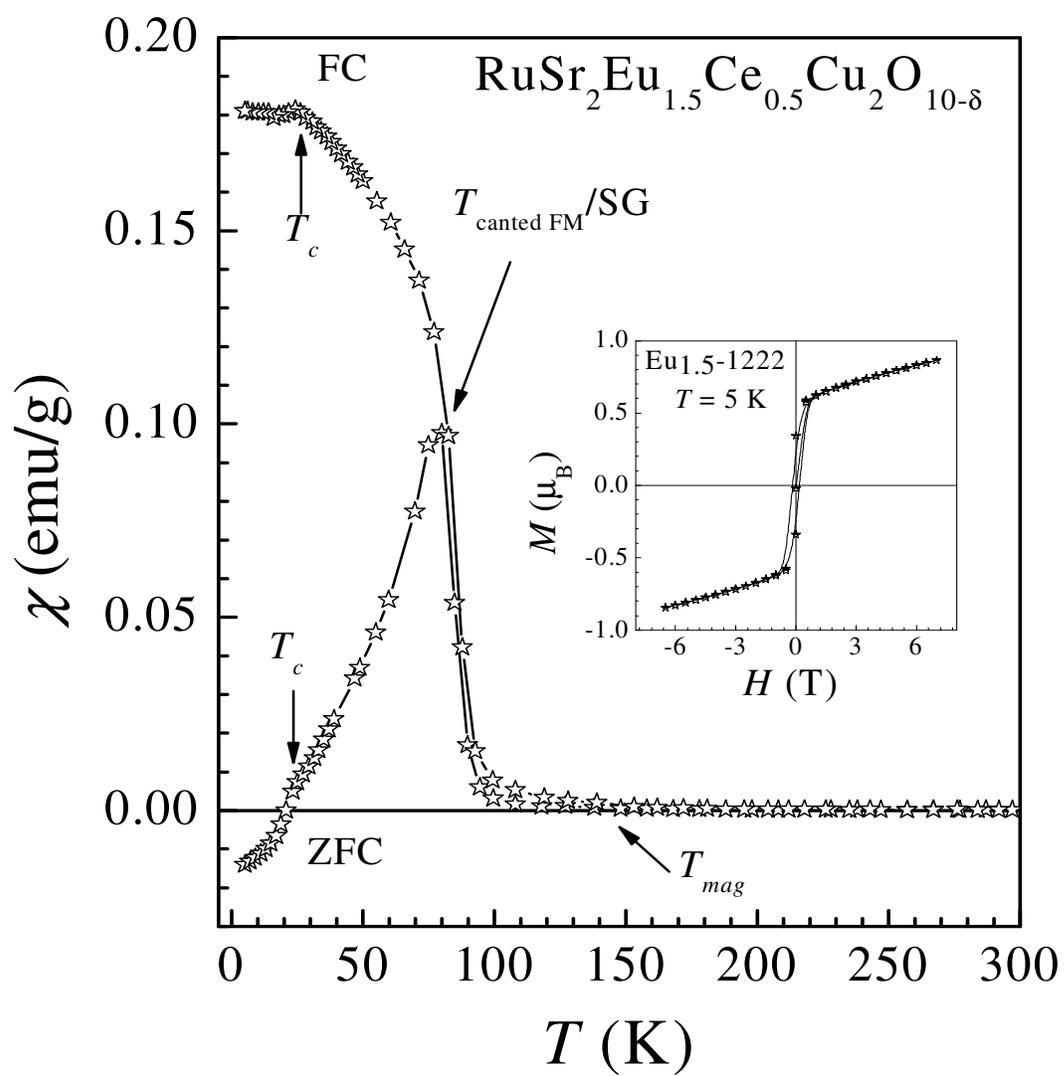

Fig. 3 R. Lal et al.

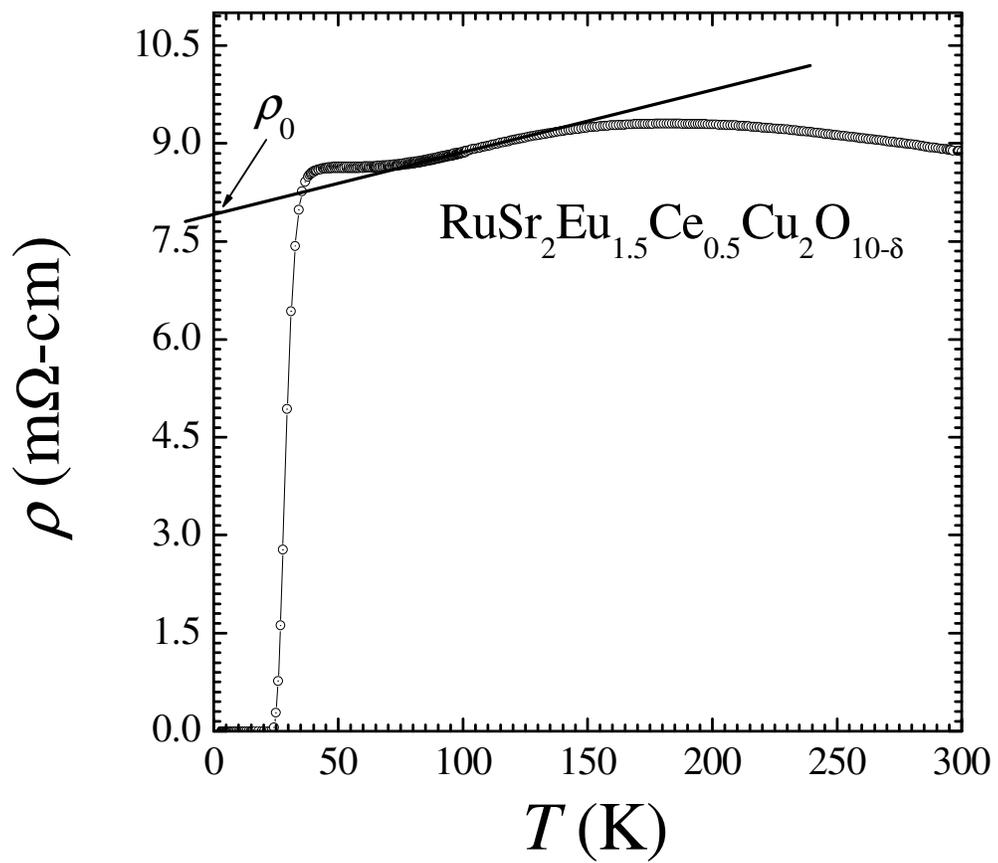

Fig. 4 R. Lal et al.

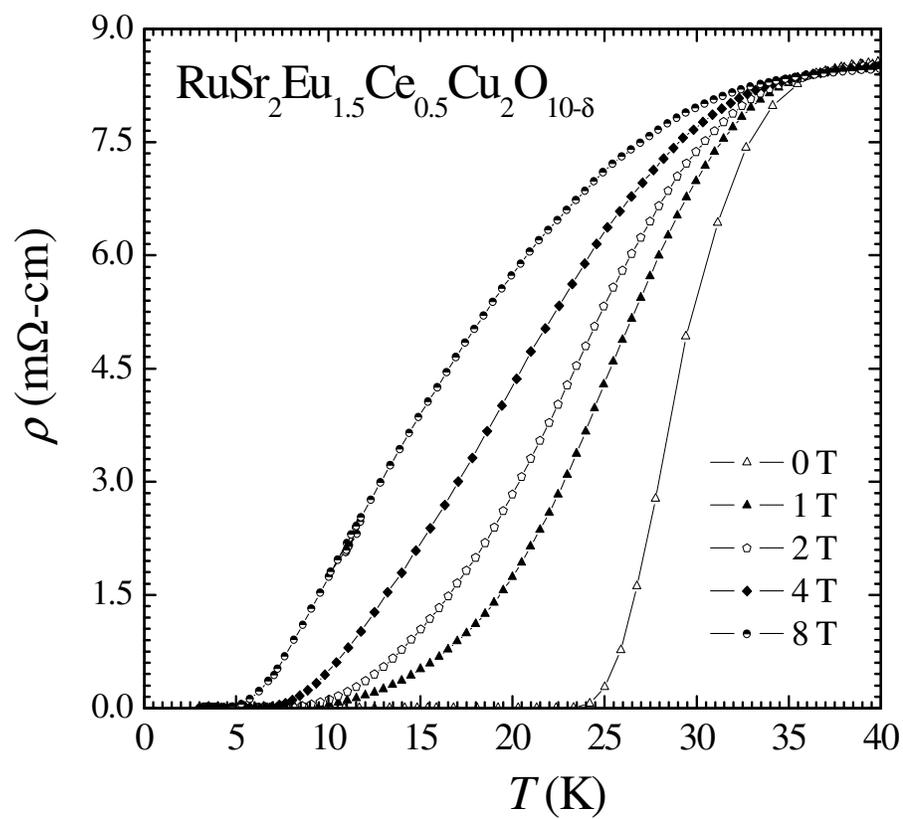

Fig. 5 R. Lal et al.

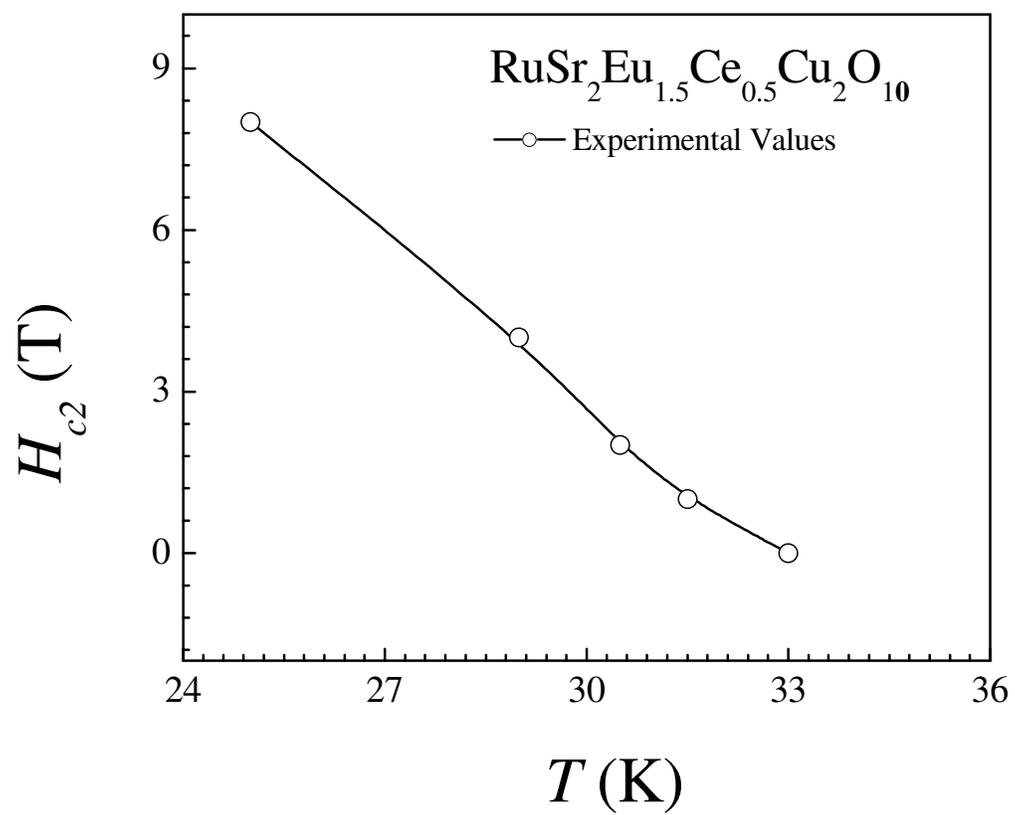

Fig. 6 R. Lal et al.

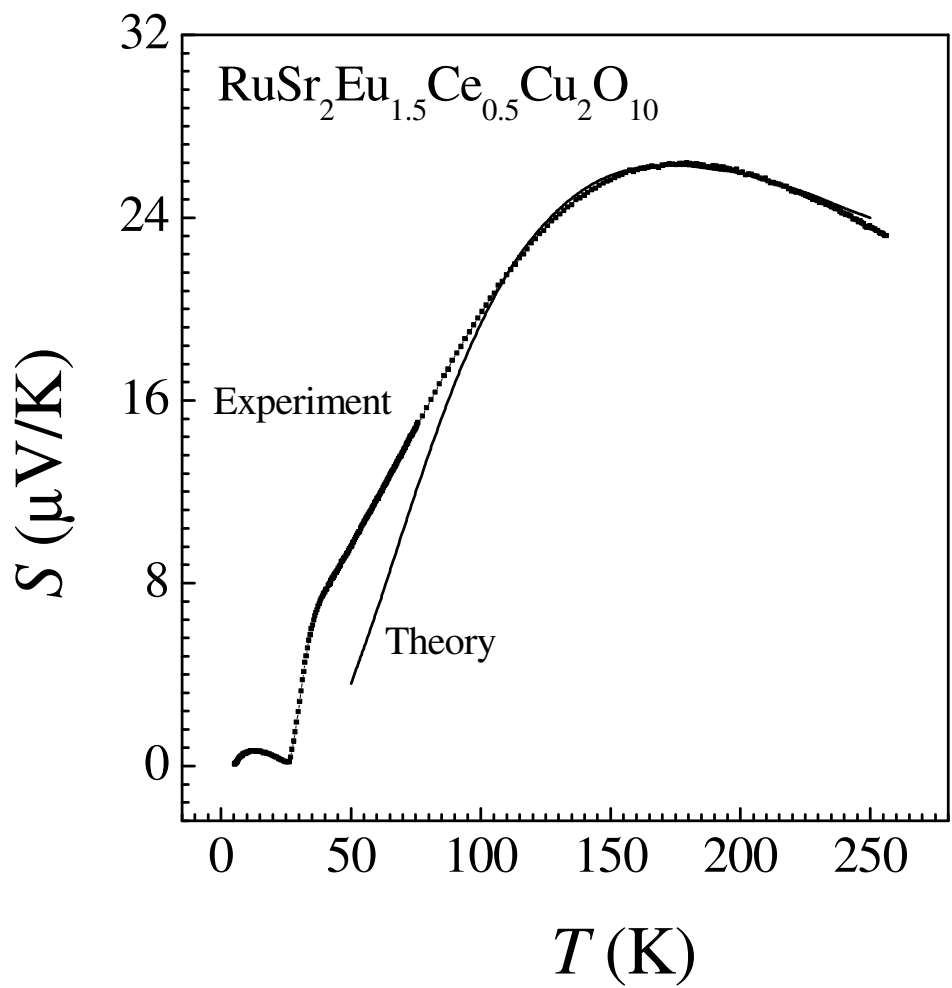